\documentstyle{aipproc}
\input boxedeps
\SetepsfEPSFSpecial %%%%% Uncomment for submission
\HideDisplacementBoxes

\newcommand{\etal}{{\em et al.}}

\newcommand{\gevcc}{\hbox{ GeV}\!/\!c^2}
\newcommand{\gev}{\hbox{ GeV}}
\newcommand{\gevc}{\hbox{ GeV}\!/\!c}

\newcommand{\mev}{\hbox{ MeV}}
\newcommand{\mevcc}{\hbox{ MeV}\!/\!c^2}
\newcommand{\tev}{\hbox{ TeV}}

\newcommand{\tevcc}{\hbox{ TeV}\!/\!c^2}

\newcommand{\mm}{\hbox{ mm}}

\newcommand{\fb}{\hbox{ fb}}

\newcommand{\m}{\hbox{ m}}

\newcommand{\eqn}[1]{(\ref{#1})}

\def\ltap{\mathop{\raisebox{-.4ex}{\rlap{$\sim$}} 
\raisebox{.4ex}{$<$}}}
\def\gtap{\mathop{\raisebox{-.4ex}{\rlap{$\sim$}} 
\raisebox{.4ex}{$>$}}}

\newcommand{\cfrac}[2]{\textstyle \frac{#1}{#2}}

\def\vev#1{\langle #1\rangle_0}
\def\abs#1{\left| #1\right|}

\def\onetev{1-TeV scale}

\newcommand{\hepex}[1]{hep-ex/#1}
\newcommand{\hepph}[1]{hep-ph/#1}
\newcommand{\hepth}[1]{hep-th/#1}
\newcommand{\nuclth}[1]{nucl-th/#1}

\newcommand{\astro}[1]{astro-ph/#1}
\def\prl#1#2#3{\frenchspacing{\it Phys. Rev. Lett. }{\bf #1}, #2 (19#3)}
\def\prll#1#2#3{\frenchspacing{\it Phys. Rev. Lett. }{\bf #1}, #2 (#3)}
\def\pr#1#2#3{\frenchspacing{\it Phys. Rev. D}{\bf #1}, #2 (19#3)}
\def\prM#1#2#3{\frenchspacing{\it Phys. Rev. D}{\bf #1}, #2 (#3)}

\def\pl#1#2#3{\frenchspacing{\it Phys. Lett. }{\bf #1}, #2 (19#3)}
\def\np#1#2#3{\frenchspacing{\it Nucl. Phys. }{\bf #1}, #2 (19#3)}
\def\rmp#1#2#3{\frenchspacing{\it Rev. Mod. Phys. }{\bf #1}, #2 (19#3)}

\def\arnps#1#2#3{\frenchspacing{\it Ann. Rev. Nucl. Part. Sci. }{\bf #1}, #2 (19#3)}
\def\ib#1#2#3{{\bf #1}, #2 (19#3)}
\def\app#1#2#3{\frenchspacing{\it Acta Phys. Polon. B}{\bf #1}, #2 (19#3)}

\def\phystoday#1#2#3#4{\frenchspacing{\it Phys. Today\/ }{\bf #1}, #2 
(\ifcase#3\or January\or 
         February\or March\or April\or May\or June\or July\or August\or 
         September\or October\or November\or December\fi, 19#4)}

\begin{document}

\title{The State of the Standard Model}
\author{Chris Quigg}
\address{Theoretical Physics Department \\ Fermi National Accelerator 
Laboratory~\thanks{Fermilab is operated by Universities Research 
Association Inc.\ under Contract No.\ DE-AC02-76CH03000 with the 
United States Department of Energy.  }\\ P.O.  Box 500, Batavia, 
Illinois 60510 USA}

\lefthead{\thepage \hfill \textsf{FERMILAB--CONF--00/021--T}}
\righthead{\textsf{FERMILAB--CONF--00/021--T}\hfill\thepage}
\maketitle
\vspace*{12pt}
\begin{center}
    \fbox{\textrm{Dedicated to the Memory of Sam Treiman}}
\end{center}
\vspace*{-12pt}

\begin{abstract}
I quickly review the successes of quantum chromodynamics.  Then I 
assess the current state of the electroweak theory, making brief 
comments about the search for the Higgs boson and some of the open 
issues for the theory.  I sketch the problems of mass and mass scales, 
and point to a speculative link between the question of identity and 
large extra dimensions.  To conclude, I return to QCD and the 
possibility that its phase structure might inform our understanding of 
electroweak symmetry breaking.
\end{abstract}

\section*{Appreciation}
Sam Treiman left us a few days ago.  With his deep respect for great 
principles, his acute understanding of field theory, and his joyful 
engagement with experiment, Sam was a model and a guide to many of us 
for whom the goal---the joy---of theoretical physics is not to gather 
up a hoard of shiny theorems, but to learn to read Nature's secrets.  
He was mentor to many of our most valued theoretical colleagues, and 
the course on particle physics he described as ``shamelessly 
phenomenological'' brought culture to generations of Princeton 
students.  He was a natural man---unpretentious and a little 
rumpled---who took wry pleasure in seeing the self-important deflated.  
A graceful writer himself\cite{pref,sbt}, he was a writer's ideal 
reader: curious, appreciative, eager to share an original idea or a 
provocative passage from his reading, and ever ready to recommend a 
good book.\footnote{His last recommendation to me was the luminous 
translations of \textit{The Iliad} and \textit{The Odyssey} by Robert 
Fagles\cite{fagles}.} Sam confessed to me, without remorse, that he 
coined ``The Standard Model,'' a curiously flat name for the marvelous 
theory of quarks and leptons that he helped to build.  I forgave him 
then, and I forgive him now---but we still need a better name!

\section*{Our Picture of Matter}
Some twenty-five years after the November Revolution, we base our 
understanding of physical phenomena on the identification of a few 
constituents that seem elementary at the current limits of resolution 
of about $10^{-18}\m$, and a few fundamental forces.  The 
constituents are the pointlike quarks
	\begin{equation}
		\left(
		\begin{array}{c}
			u  \\
			d
		\end{array}
		 \right)_{L} \;\;\;\;\;\;
		\left(
		\begin{array}{c}
			c  \\
			s
		\end{array}
		 \right)_{L} \;\;\;\;\;\;
		\left(
		\begin{array}{c}
			t  \\
			b
		\end{array}
		 \right)_{L}		 
	\end{equation}
and leptons
		\begin{equation}
		\left(
		\begin{array}{c}
			\nu_{e}  \\
			e^{-}
		\end{array}
		 \right)_{L} \;\;\;\;\;\;
		\left(
		\begin{array}{c}
			\nu_{\mu}  \\
			\mu^{-}
		\end{array}
		 \right)_{L} \;\;\;\;\;\;
		\left(
		\begin{array}{c}
			\nu_{\tau}  \\
			\tau^{-}
		\end{array}
		 \right)_{L}		 
	\end{equation}
with strong, weak, and electromagnetic interactions specified by
$SU(3)_{c}\otimes SU(2)_{L}\otimes U(1)_{Y}$ gauge symmetries.

This concise statement of the standard model invites us to consider 
the agenda of particle physics today under four themes.  
\textit{Elementarity.} Are the quarks and leptons structureless, or 
will we find that they are composite particles with internal 
structures that help us understand the properties of the individual 
quarks and leptons?  \textit{Symmetry.} One of the most powerful 
lessons of the modern synthesis of particle physics is that 
symmetries prescribe interactions.  Our investigation of symmetry must 
address the question of which gauge symmetries exist (and, eventually, 
why).  We must also understand how the electroweak 
symmetry\footnote{and, no doubt, others---including the symmetry that 
brings together the strong, weak, and electromagnetic interactions.} 
is hidden.  The most urgent problem in particle physics is to complete 
our understanding of electroweak symmetry breaking by exploring the 
1-TeV scale.  \textit{Unity.} We have the fascinating possibility of 
gauge coupling unification, the idea that all the interactions we 
encounter have a common origin and thus a common strength at suitably 
high energy.  Next comes the imperative of anomaly freedom in the 
electroweak theory, which urges us to treat quarks and leptons 
together, not as completely independent species.  Both ideas are 
embodied in unified theories of the strong, weak, and electromagnetic 
interactions, which imply the existence of still other forces---to 
complete the grander gauge group of the unified theory---including 
interactions that change quarks into leptons.  The self-interacting 
quanta of non-Abelian theories and supersymmetry both hint that the 
traditional distinction between force particles and constituents might 
give way to a unified understanding of all the particles.  
\textit{Identity.} We do not understand the physics that sets quark 
masses and mixings.  Although experiments are testing the idea that the phase 
in the quark-mixing matrix lies behind the observed \textsf{CP} 
violation, we do not know what determines that phase.  The 
accumulating evidence for neutrino oscillations presents us with a new 
embodiment of these puzzles in the lepton sector.  At bottom, the 
question of identity is very simple to state: What makes an electron 
an electron, a neutrino a neutrino, and a top quark a top quark?

\section*{QCD Is Part of the Standard Model}
The quark model of hadron structure and the parton model of 
hard-scattering processes have such pervasive influence on the way we 
conceptualize particle physics that quantum chromodynamics, the theory 
of strong interactions that underlies both, often fades into the 
background when the standard model is discussed.  I want to begin my 
state of the standard model report with the clear statement that QCD 
is indeed part of the standard model, and with the belief that 
understanding QCD may be indispensable for deepening our understanding 
of the electroweak theory.

Quantum chromodynamics is a remarkably simple, successful, and rich 
theory of the strong interactions.\footnote{For a passionate elaboration 
of this statement, see Frank Wilczek's keynote address at PANIC '99, 
Ref.\ \cite{fw}.  An authoritative portrait of QCD and its many applications 
appears in the monograph by Ellis, Stirling, and Webber, Ref.\ \cite{esw}.}  The 
perturbative regime of QCD exists, thanks to the crucial property of 
asymptotic freedom, and describes many phenomena in quantitative 
detail.  The strong-coupling regime controls hadron structure and 
gives us our best information about quark masses.

\begin{figure}[tb] 
\centerline{\BoxedEPSF{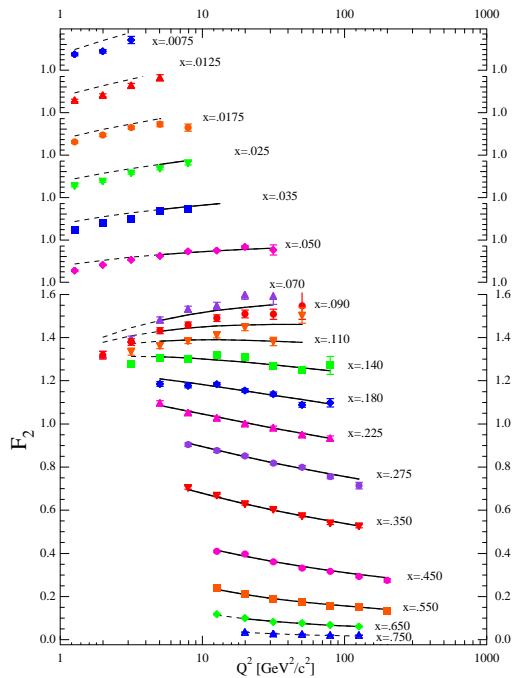 scaled 470}}
\vspace{10pt}
\caption{The structure function $F_2$ measured in $\nu N$ interactions,
%charged-current neutrino-nucleon interactions.  The errors shown are 
%statistical; systematic uncertainties are approximately $\pm 2\%$.  
%The lines are the result of a QCD fit to the data.  
from Ref.\ {\protect\cite{csb}}.}
\label{fig:F2}
\end{figure}

The classic test of perturbative QCD is the prediction of subtle 
violations of Bjorken scaling in deeply inelastic lepton scattering.  
As an illustration of the current state of the comparison between 
theory and experiment, I show in Figure \ref{fig:F2} the singlet 
structure function $F_{2}(x,Q^{2})$ measured in $\nu N$ 
charged-current interactions by the CCFR Collaboration at Fermilab.  
The solid lines for $Q^{2} \gtap (5\gevc)^{2}$ represent QCD fits; 
the dashed lines are extrapolations to smaller values of $Q^{2}$.  As 
we see in this example, modern data are so precise that one can search 
for small departures from the QCD expectation.

\begin{figure}[tb] 
\centerline{\BoxedEPSF{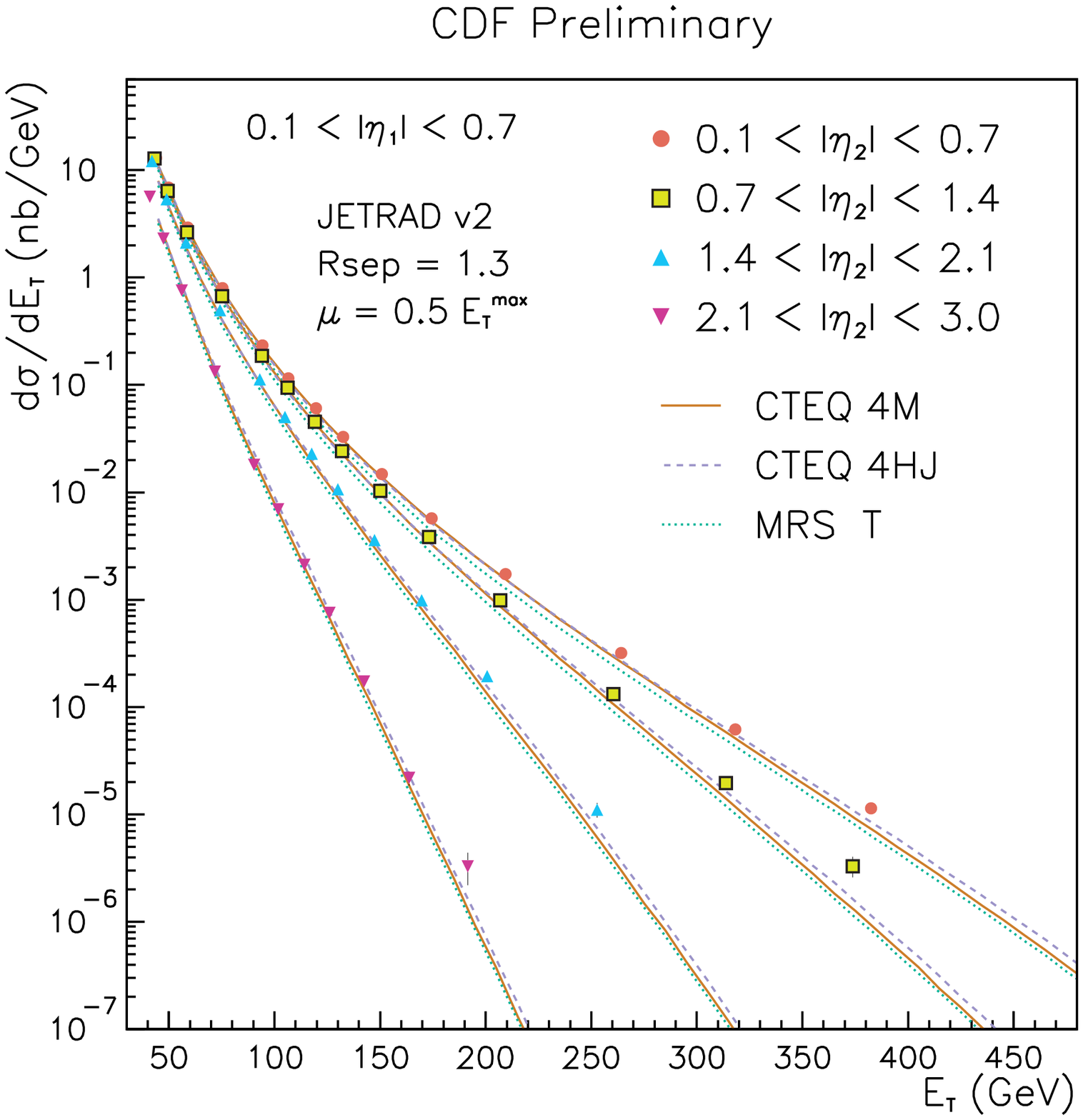 scaled 400}}
\vspace{10pt}
\caption{Cross sections measured at $\sqrt{s} = 1.8\tev$ by the CDF 
Collaboration for central jets (defined by $0.1 < \abs{\eta_{1}} < 
0.7$), with the second jet confined to specified intervals in the 
pseudorapidity $\eta_{2}$ {\protect\cite{cdfjets}}.  The curves show 
next-to-leading-order QCD predictions based on the CTEQ4M (solid 
line), CTEQ4HJ (dashed line), and MRST (dotted line) parton 
distributions.}
\label{fig:jets}
\end{figure}

Perturbative QCD also makes spectacularly successful predictions for 
hadronic processes.  I show in Figure \ref{fig:jets} that pQCD, 
evaluated at next-to-leading order using the program \textsc{jetrad,} 
accounts for the transverse-energy spectrum of central jets produced 
in the reaction
\begin{equation}
    \bar{p}p \to \hbox{jet}_{1} + \hbox{jet}_{2} + \hbox{anything}
    \label{eq:pbarpjj}
\end{equation}
over at least six orders of magnitude, at $\sqrt{s} = 
1.8\tev$. \footnote{For a systematic review of high-$E_{T}$ jet 
production, see Blazey and Flaugher, Ref.\ \cite{brenna}.}

%%%%%%%%%%%%%%%%%%%%%%%%%%%%%%%%%%%%%%%%%%%%%%%%%%%%%%%%%%%%%%%%%%%%%%
%                                                                    %
%   The running of the strong coupling constant in CDF jets \ldots   %
%   Figure \ref{fig:CDFalpha} \ldots                                 %
%   \begin{figure}[tb]                                               %
%   \centerline{\BoxedEPSF{CDFRunningalpha.eps scaled 600}}          %
%   \vspace{10pt}                                                    %
%   \caption{Running $\alpha_{s}$ from CDF \ldots}                   %
%   \label{fig:CDFalpha}                                             %
%   \end{figure}                                                     %
%                                                                    %
%%%%%%%%%%%%%%%%%%%%%%%%%%%%%%%%%%%%%%%%%%%%%%%%%%%%%%%%%%%%%%%%%%%%%%

\begin{figure}[tb] 
\centerline{\BoxedEPSF{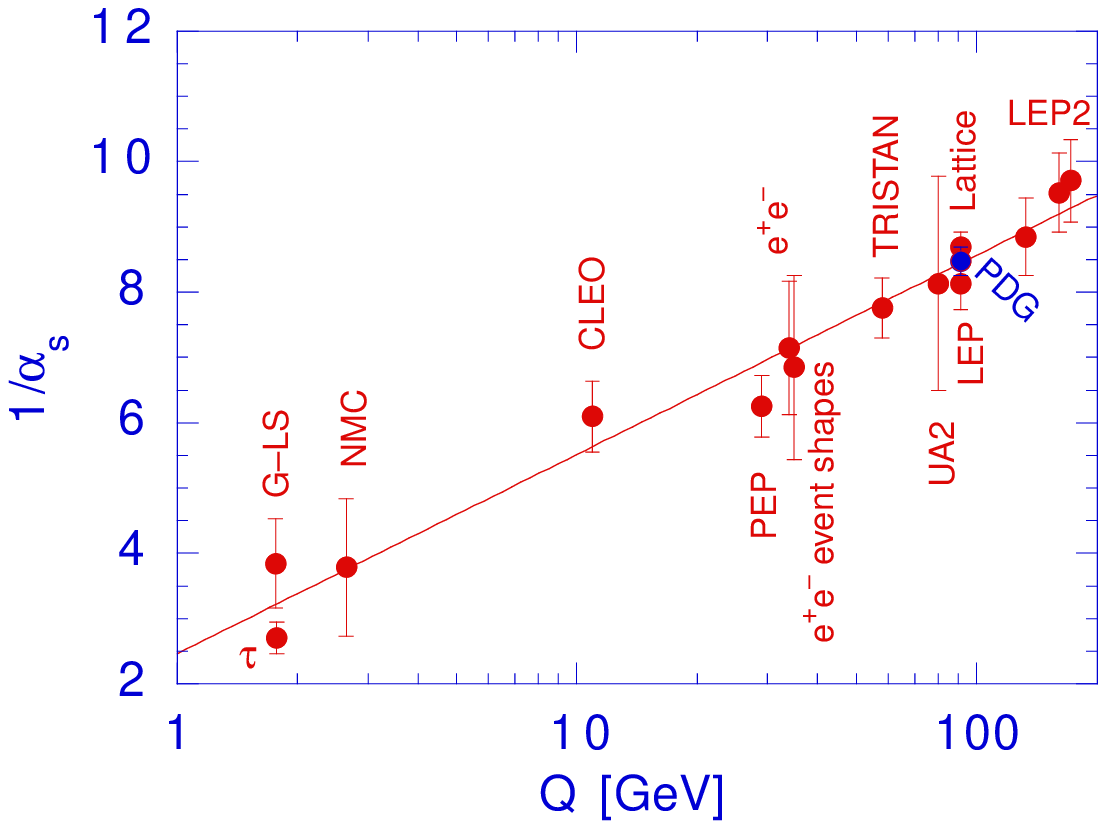 scaled 1000}}
\vspace{10pt}
\caption{Determinations of $1/\alpha_{s}$, plotted at the scale 
$\mu$ at which the measurements were made.  The line shows the 
expected evolution {\protect\eqn{eq:runalph}}.}
\label{fig:oneoveralpha}
\end{figure}

\begin{figure}[tb] 
\centerline{\BoxedEPSF{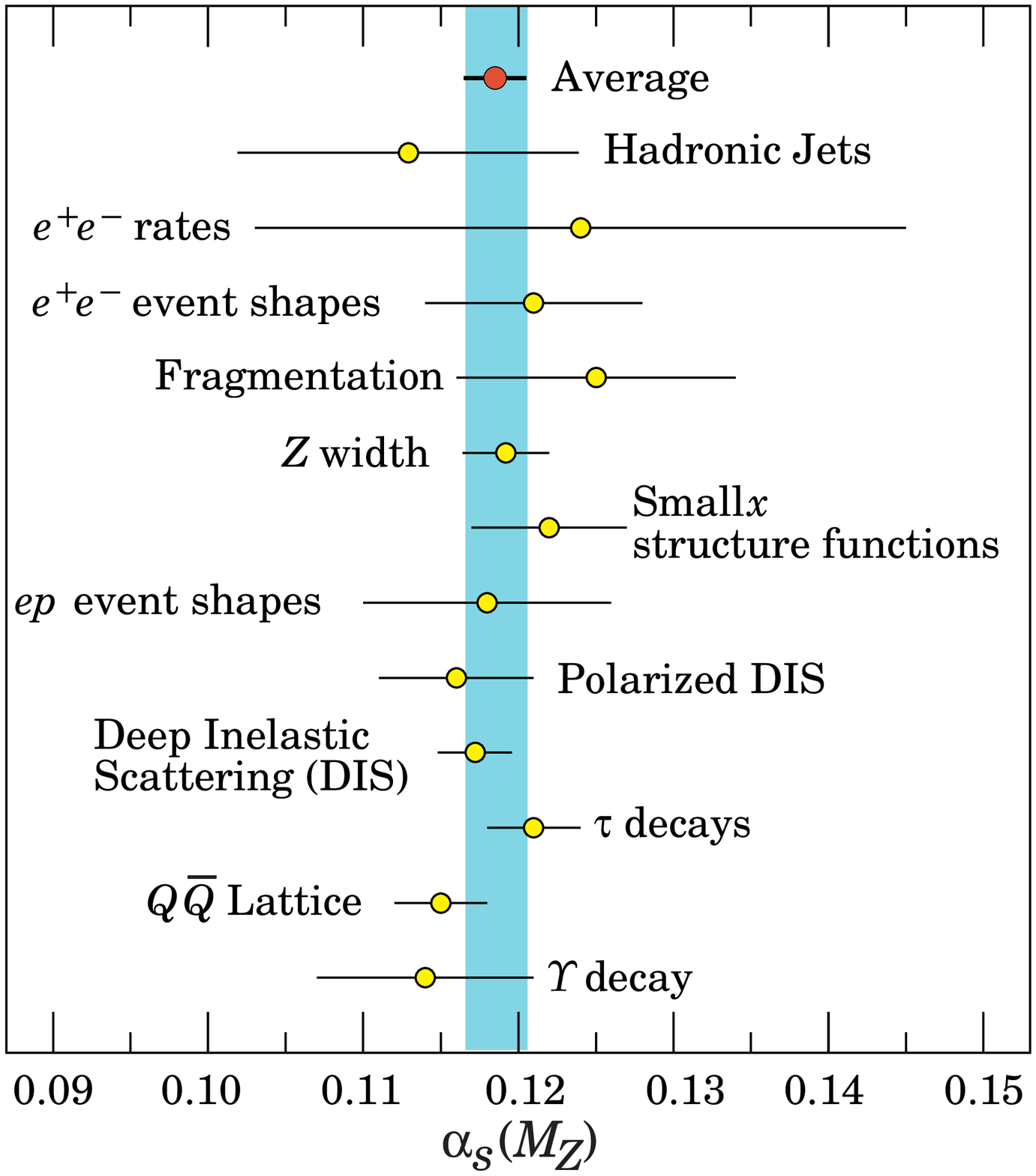 scaled 450}}
\vspace{10pt}
\caption{Determinations of $\alpha_{s}(M_{Z})$ from several 
processes. In most cases, the value measured at a scale $\mu$ has been 
evolved to $\mu = M_{Z}$.  Error bars include the theoretical uncertainties.
From the \textit{Review of Particle Physics} {\protect\cite{pdg}}.}
\label{fig:asMZ}
\end{figure}

The $Q^{2}$-evolution of the strong coupling constant predicted by 
QCD, which in lowest order is
\begin{equation}
    1/\alpha_{s}(Q^{2}) = 1/\alpha_{s}(\mu^{2}) +
    \frac{33-2n_{f}}{12\pi} \log(Q^{2}/\mu^{2}) \; ,
    \label{eq:runalph}
\end{equation}
where $n_{f}$ is the number of active quark flavors, has been 
observed within individual experiments \cite{cdfrun,leprun} and by 
comparing determinations made in different experiments at different 
scales.  A compilation of $1/\alpha_{s}$ determinations from many 
experiments, shown in Figure \ref{fig:oneoveralpha}, exhibits the 
expected behavior.\footnote{A useful plot of $\alpha_{s}$ 
\textit{vs.} $Q^{2}$ appears as Figure 9.2 of the \textit{Review of Particle 
Physics,} Ref.\ \cite{pdg}.}
%%%%%%%%%%%%%%%%%%%%%%%%%%%%%%%%%%%%%%%%%%%%%%%%%%%%%%%%%%%%%%%%%%%%%%%%%%%%%%%
%                                                                             %
%   Evolution of $\alpha_{s}$ vs. $Q^{2}$ is Figure \ref{fig:Ralpha}          %
%   \ldots \cite{pdg}                                                         %
%   \begin{figure}[tb]                                                        %
%   \centerline{\BoxedEPSF{qcdFig92.eps scaled 600}}                          %
%   \vspace{10pt}                                                             %
%   \caption{Determinations of $\alpha_{s}(\mu)$, plotted at the scale        %
%   $\mu$ at which the measurements were made.  The line shows the central    %
%   values and $ \pm 1\sigma$ boundaries of the Particle Data Group's         %
%   average, from Ref.\ {\protect\cite{pdg}}.}                                %
%   \label{fig:Ralpha}                                                        %
%   \end{figure}                                                              %
%                                                                             %
%%%%%%%%%%%%%%%%%%%%%%%%%%%%%%%%%%%%%%%%%%%%%%%%%%%%%%%%%%%%%%%%%%%%%%%%%%%%%%%
When evolved to a common scale $\mu = M_{Z}$, the various 
determinations of $\alpha_{s}$ lead to consistent values, as shown in 
Figure \ref{fig:asMZ}.

\begin{figure}[tb] 
\centerline{\BoxedEPSF{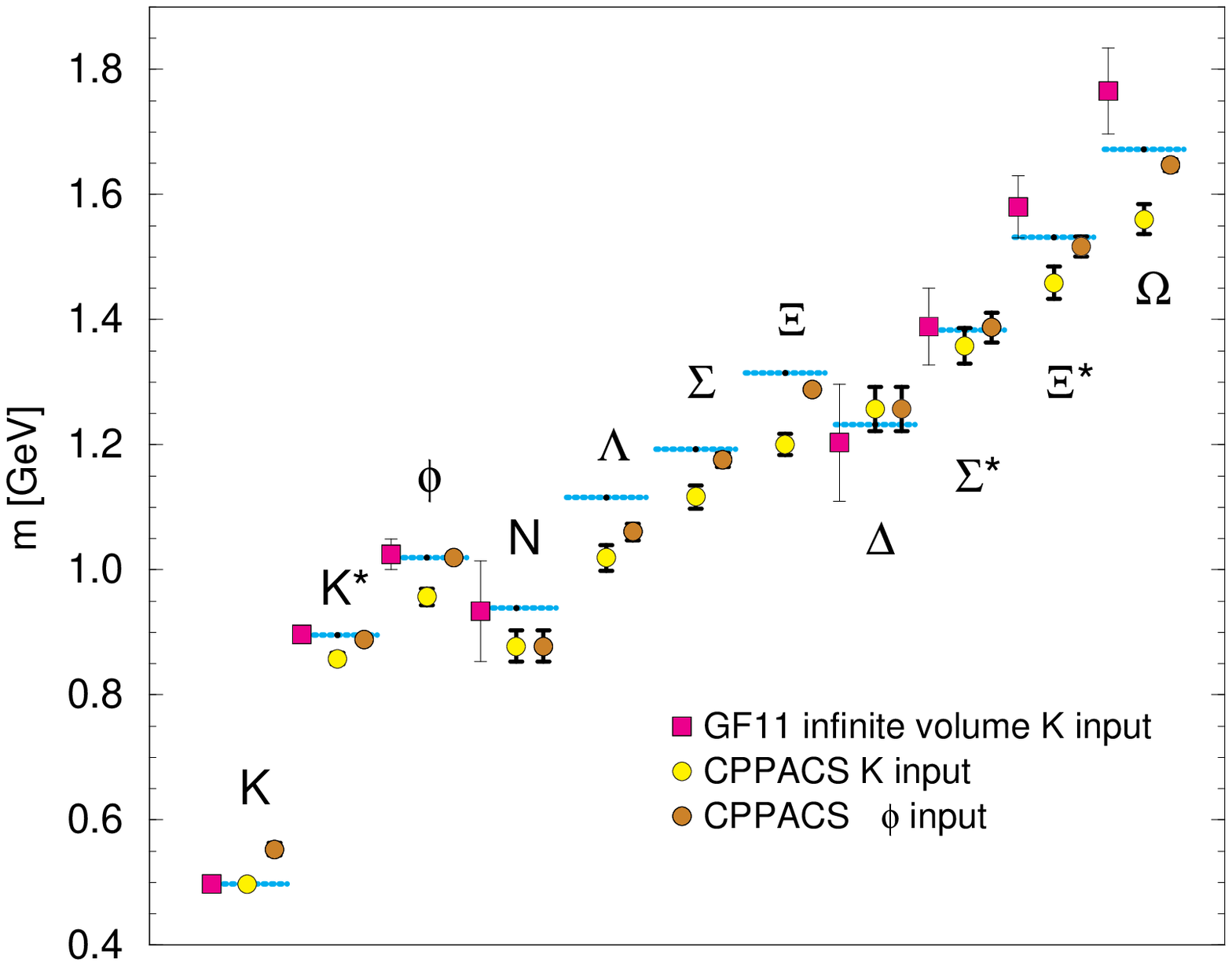 scaled 600}}
\vspace{10pt}
\caption{Final results of the CP-PACS Collaboration's quenched light 
hadron spectrum in the continuum limit {\protect\cite{burk}}.  
Experimental values (horizontal lines) and earlier results from the 
GF11 Collaboration {\protect\cite{ref:GF11}} are plotted for 
comparison.}
\label{fig:hadrons}
\end{figure}

Thanks to QCD, we have learned that the dominant contribution to the 
light-hadron masses is not the masses of the quarks of which they are 
constituted, but the energy stored up in confining the quarks in a 
tiny volume.\footnote{An accessible essay on our understanding of hadron 
mass appears in Ref.\ \cite{fwpt}.} Our most useful tool in the 
strong-coupling regime is lattice QCD.  Calculating the light hadron 
spectrum from first principles has been one of the main objectives of 
the lattice program, and important strides have been made recently.  
In 1994, the GF11 Collaboration \cite{ref:GF11}carried out a quenched 
calculation of the spectrum (no dynamical fermions) that yielded 
masses that agree with experiment within 5--10\%, with good 
understanding of the residual systematic uncertainties.  The CP-PACS 
Collaboration centered in Tsukuba has embarked on an ambitious program 
that will soon lead to a full (unquenched) calculation.  Their 
quenched results, along with those of the GF11 Collaboration, are 
presented in Figure \ref{fig:hadrons} \cite{burk}.  The gross features 
of the light-hadron spectrum are reproduced very well, but if you look 
with a critical eye (as the CP-PACS collaborators do), you will notice
that the quenched light hadron spectrum systematically deviates from 
experiment.  The $K$-$K^*$ mass splitting is underestimated by about 
$10\%$, and the results differ depending on whether the 
strange-quark mass is fixed from the $K$ mass or the $\phi$ mass.  The 
forthcoming unquenched results should improve the situation further, 
and give us new insights into how well---and why!---the simple quark 
model works.
\clearpage

\section*{The $SU(2)_{L}\otimes U(1)_{Y}$ Electroweak Theory}
\begin{figure}[tbh]
	\centerline{\BoxedEPSF{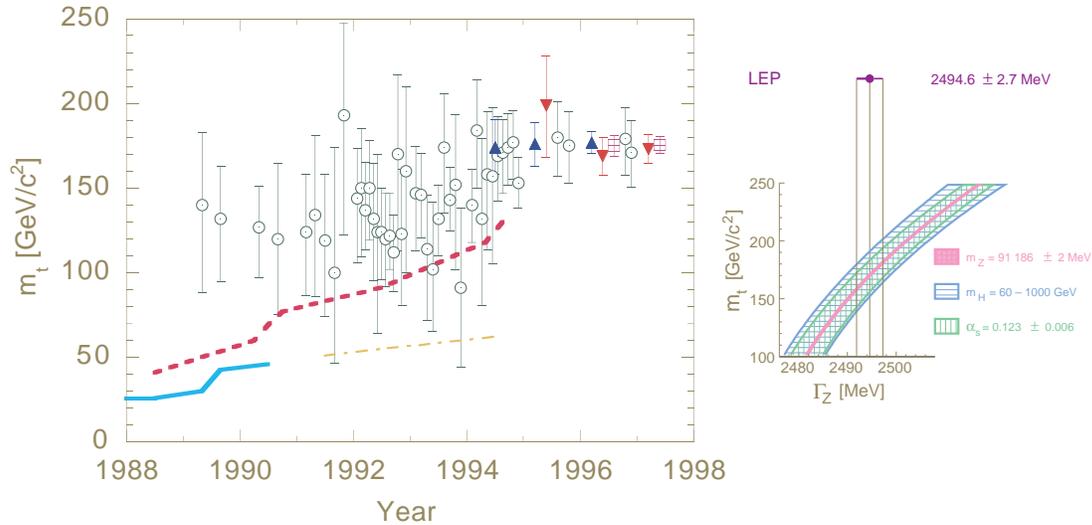  scaled 550}}
\caption{Indirect determinations of the top-quark mass from fits to 
electroweak observables (open circles) and 95\% confidence-level lower 
bounds on the top-quark mass inferred from direct searches in 
$e^{+}e^{-}$ annihilations (solid line) and in $\bar{p}p$ collisions, 
assuming that standard decay modes dominate (broken line).  An 
indirect lower bound, derived from the $W$-boson width inferred from 
$\bar{p}p \rightarrow (W\hbox{ or }Z)+\hbox{ anything}$, is shown as 
the dot-dashed line.  Direct measurements of $m_{t}$ by the CDF 
(triangles) and D\O\ (inverted triangles) Collaborations are shown at 
the time of initial evidence, discovery claim, and at the conclusion 
of Run 1.  The world average from direct observations is shown as the
crossed box.  For sources of data, see Ref.  {\protect\cite{pdg}}. 
\textit{Inset:} Electroweak theory predictions for the width of the
$Z^{0}$ boson as a function of the top-quark mass, compared with the
width measured in LEP experiments.  (From Ref.\
{\protect\cite{cqpt}}).}
	\label{EWtop}
\end{figure}

The electroweak theory is founded on the weak-isospin symmetry 
embodied in the doublets
\begin{equation}
 		\left(
		\begin{array}{c}
			u  \\
			d^{\prime}
		\end{array}
		 \right)_{L} \;\;
		\left(
		\begin{array}{c}
			c  \\
			s^{\prime}
		\end{array}
		 \right)_{L} \;\;
		\left(
		\begin{array}{c}
			t  \\
			b^{\prime}
		\end{array}
		 \right)_{L} \;\;\;\;\;\;
		\left(
		\begin{array}{c}
			\nu_{e}  \\
			e^{-}
		\end{array}
		 \right)_{L} \;\;
		\left(
		\begin{array}{c}
			\nu_{\mu}  \\
			\mu^{-}
		\end{array}
		 \right)_{L} \;\;
		\left(
		\begin{array}{c}
			\nu_{\tau}  \\
			\tau^{-}
		\end{array}
		 \right)_{L}
    \label{eq:ewth}
\end{equation}
and weak-hypercharge phase symmetry, plus the idealization that 
neutrinos are massless.\footnote{For a survey of the electroweak 
theory, with many references, see Ref.\ \cite{granada}.} In its 
simplest form, with the electroweak gauge symmetry broken by the Higgs 
mechanism, the $SU(2)_{L}\otimes U(1)_{Y}$ theory has scored many 
qualitative successes: the prediction of neutral-current interactions, 
the necessity of charm, the prediction of the existence and properties 
of the weak bosons $W^{\pm}$ and $Z^{0}$.  Over the past ten years, in 
great measure due to the beautiful experiments carried out at the $Z$ 
factories at CERN and SLAC, precision measurements have tested the 
electroweak theory as a quantum field theory, at the one-per-mille 
level \cite{sirlin,morris}.  As an example of the insights precision 
measurements have brought us (one that mightily impressed the Royal 
Swedish Academy of Sciences), I show in Figure \ref{EWtop} the time 
evolution of the top-quark mass favored by simultaneous fits to many 
electroweak observables.  Higher-order processes involving virtual top 
quarks are an important element in quantum corrections to the 
predictions the electroweak theory makes for many observables.  A case 
in point is the total decay rate, or width, of the $Z^{0}$ boson: the 
comparison of experiment and theory shown in the inset to Figure 
\ref{EWtop} favors a top mass in the neighborhood of $180\gevcc$.

\begin{figure}[tb] 
\centerline{\BoxedEPSF{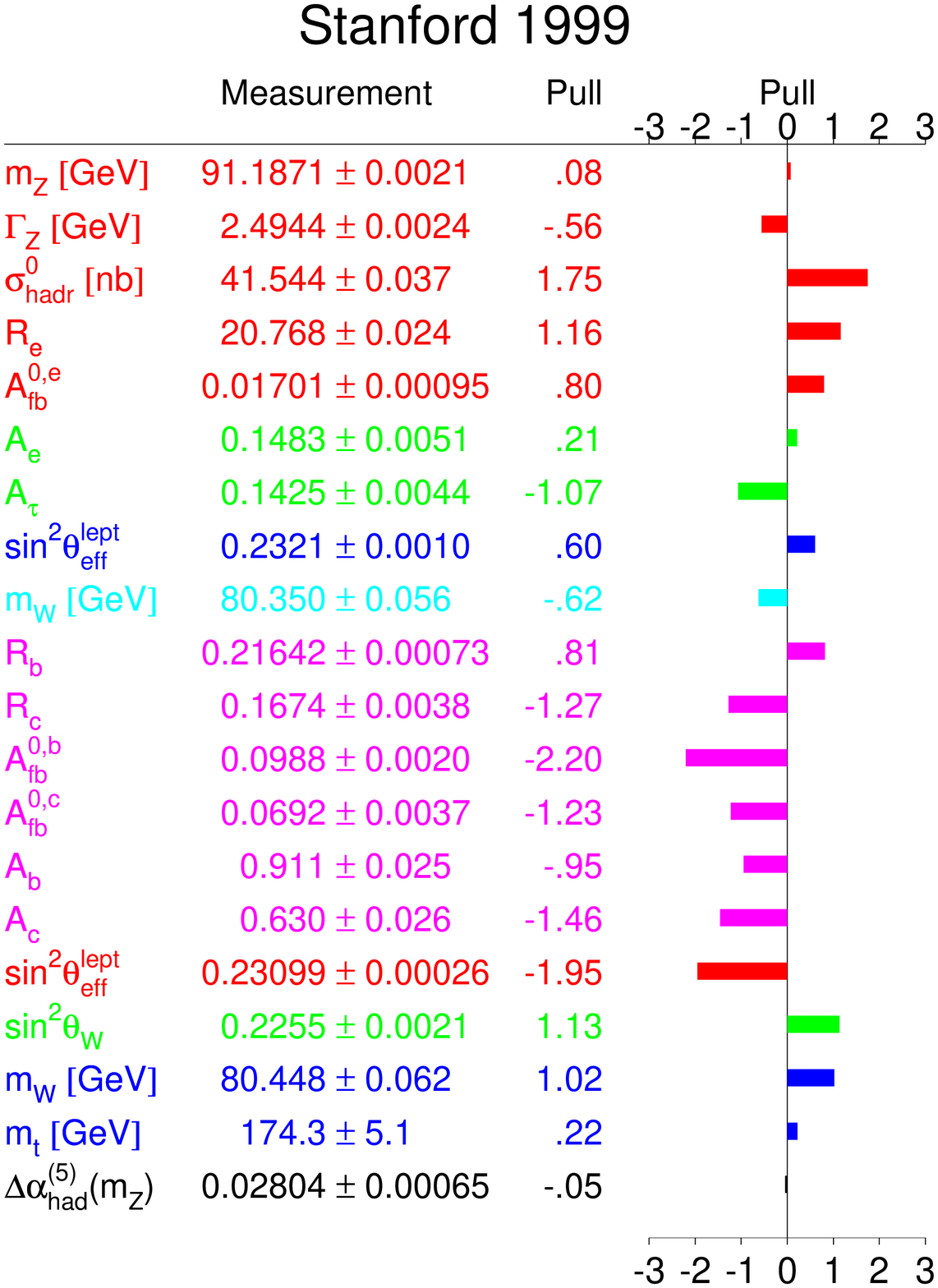 scaled 400}}
\vspace{10pt}
\caption{Precision electroweak measurements and the pulls they exert 
on a global fit to the standard model, from Ref.\ 
{\protect\cite{ewwg}}.}
\label{fig:pulls}
\end{figure}

The comparison between the electroweak theory and a considerable 
universe of data is shown in Figure \ref{fig:pulls} \cite{ewwg}, 
where the pull, or difference between the global fit and measured 
value in units of standard deviations, is shown for some twenty 
observables.
The distribution of pulls for this fit, due to the LEP Electroweak 
Working Group, is not noticeably different from a normal 
distribution, and only a couple of observables differ from the fit by 
as much as about two standard deviations.  This is the case for any 
of the recent fits, including two that I point to later as examples of 
what we may learn from global fits about physics beyond the standard 
model \cite{kaoru,erlang}.  From fits of the kind represented here, 
we learn that the standard-model interpretation of the data favors a 
light Higgs boson.  We will hear about these conclusions in more 
detail from Bill Marciano \cite{wjmsf}, and I will add some cautionary 
remarks shortly.

The beautiful agreement between the electroweak theory and a vast 
array of data from neutrino interactions, hadron collisions, and 
electron-positron annihilations at the $Z^{0}$ pole and beyond means 
that electroweak studies have become a modern arena in which we can 
look for new physics ``in the sixth place of 
decimals.''\footnote{In \textit{The Odd Quantum,} Sam 
Treiman\cite{sbt} quotes from the 1898--99 University of Chicago 
catalogue: ``While it is never safe to affirm that the future of the 
Physical Sciences has no marvels in store even more astonishing than 
those of the past, it seems probable that most of the grand underlying 
principles have been firmly established and that further advances are 
to be sought chiefly in the rigorous application of these principles 
to all the phenomena which come under our notice \ldots .  An eminent 
physicist has remarked that the future truths of Physical Science are 
to be looked for in the sixth place of decimals.'' Future truths are 
to be found still in precision measurements, but the century we are 
leaving has repeatedly shown that Nature's marvels are not 
limited by our imagination, and that exploration can yield surprises 
that completely change the way we think.}

 % refer to W.  J.  Marciano, \cite{Marciano:1999ia}.

%%%%%%%%%%%%%%%%%%%%%%%%%%%%%%%%%%%%%%%%%%%%%%%%%%%%%%%%%%%%%%%%%%%%%%%%%%%%%%%%%
%                                                                               %
%   (Omit LEPEWWG plot of $M_{W}$ \textit{vs.} $m_{t}$ and Higgs mass \ldots)   %
%   \begin{figure}[tb]                                                          %
%   \centerline{\BoxedEPSF{t99_mt_mw_contours.eps scaled 600}}                  %
%   \vspace{10pt}                                                               %
%   \caption{$M_{W},m_{t}$ contours \ldots}                                     %
%   \label{fig:contours}                                                        %
%   \end{figure}                                                                %
%                                                                               %
%%%%%%%%%%%%%%%%%%%%%%%%%%%%%%%%%%%%%%%%%%%%%%%%%%%%%%%%%%%%%%%%%%%%%%%%%%%%%%%%%

\subsection*{The Higgs Boson Search at the Tevatron}
The most promising channel for Higgs-boson searches at the Tevatron will be the 
$b\bar{b}$ mode, for which the branching fraction exceeds about 50\% 
throughout the region preferred by supersymmetry and the precision 
electroweak data.  At the Tevatron, the direct production of a light 
Higgs boson in gluon-gluon fusion $gg \rightarrow H \rightarrow 
b\bar{b}$ is swamped by the ordinary QCD production of $b\bar{b}$ 
pairs.  The high background in the $b\bar{b}$ channel means that 
special topologies must be employed to improve the ratio of signal to 
background and the significance of an observation.  The high 
luminosities that can be contemplated for a future run argue that the 
associated-production reactions $\bar{p}p \to H(W,Z) + \hbox{anything}$
are plausible candidates for a Higgs discovery at the Tevatron.

The prospects for exploiting these reactions were explored in detail 
in connection with the Run 2 Supersymmetry / Higgs Workshop at 
Fermilab,\footnote{Work carried out in the context of the Tevatron Run
2 Supersymmetry/Higgs Workshop at Fermilab may be found at
\textsf{http://fnth37.fnal.gov/susy.html}.} and will be presented in 
detail at this meeting by John Hobbs \cite{hobbs}. Taking into account what
is known, and what might conservatively be expected, about
sensitivity, mass resolutions, and background rejection, these
investigations show that it is unlikely that a standard-model Higgs 
boson could be observed in Tevatron Run 2.  The prospects are much 
brighter for Run 2$^{\mathit{bis}}$.  Indeed, the sensitivity to a 
light Higgs boson is what motivates the integrated luminosity of 
$30\fb^{-1}$ specified for Run 2$^{\mathit{bis}}$.

\begin{figure}[tb] 
\centerline{\BoxedEPSF{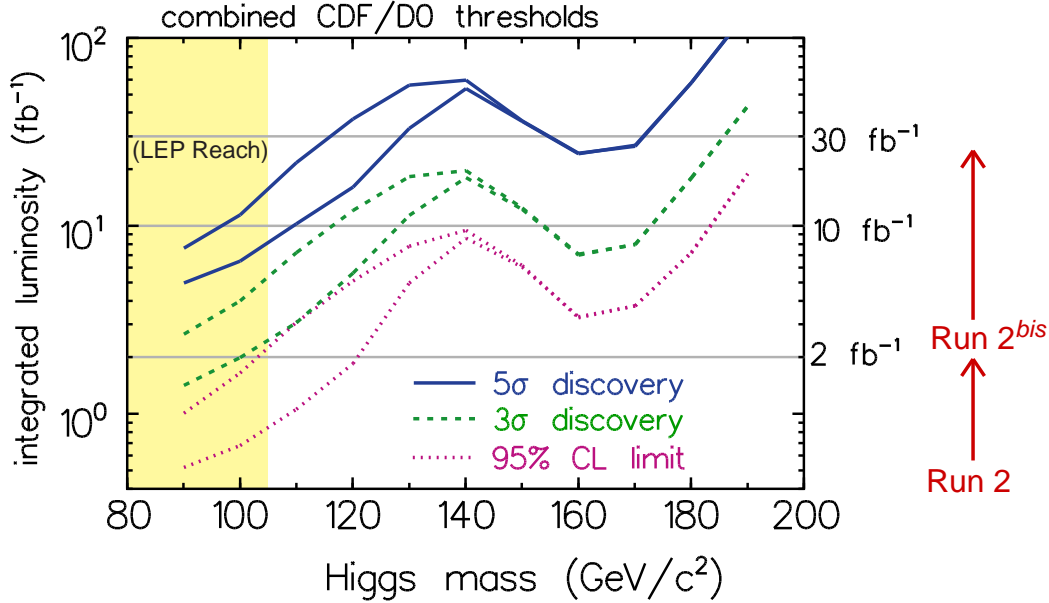 scaled 750}}
\vspace{10pt}
\caption{Integrated luminosity projected for the detection of a 
standard-model Higgs boson at the Tevatron Collider.  In each pair of 
curves, the lower reach corresponds to a generic analysis, the upper 
reach to a neural-net optimization.  See the talk by John Hobbs 
{\protect\cite{hobbs}} for a detailed exposition of the assumptions.}
\label{fig:Run2H}
\end{figure}

The detection strategy evolved in the Supersymmetry / Higgs Workshop 
involves combining the $HZ$ and $HW$ signatures and adding the data 
from the CDF and D\O\ detectors.  Apart from the increase in 
luminosity, the key ingredient in the heightened sensitivity is a 
projected improvement in the $b\bar{b}$ mass resolution to $10\%$.  
Prospects are summarized in Figure \ref{fig:Run2H}, which shows as a 
function of the Higgs-boson mass the luminosity required for exclusion 
at 95\%\ confidence level (dotted lines), three-standard-deviation 
evidence (dashed lines), and five-standard-deviation discovery (solid 
lines).  We see that an integrated luminosity of $2\fb^{-1}$, expected 
in Run 2, is insufficient for a convincing observation of a 
standard-model Higgs boson with a mass too large to be observed at 
LEP~2.  However, a 95\%\ CL exclusion is possible up to about 
$125\gevcc$.  On the other hand, about $10\fb^{-1}$ would permit 
detailed study of a standard-model Higgs boson discovered at LEP~2.  
If the Higgs boson lies beyond the reach of LEP~2, $M_{H} \gtap (105 
\hbox{ -- }110)\gevcc$, then a 5-$\sigma$ discovery will be possible 
in a future Run 2$^{\mathit{bis}}$ of the Tevatron ($30\fb^{-1}$) for 
masses up to about $(125\hbox{ -- }130)\gevcc$.  This prospect is the 
most powerful incentive we have for Run 2$^{\mathit{bis}}$.  Over the 
range of masses accessible in associated production at the Tevatron, 
it should be possible to determine the mass of the Higgs boson to $\pm 
(1\hbox{ -- }3)\gevcc$.

Recent studies \cite{hantz} suggest that it may be possible to extend 
the reach of the Tevatron significantly by making use of the 
real-$W$--virtual-$W$ ($WW^{*}$) decay modes for Higgs boson produced 
in the elementary reaction $gg \to H$.  The $WW^{*}$ channel has the largest 
branching fraction for $M_{H} \gtap 140\gevcc$.  According to the 
analysis summarized in Figure \ref{fig:Run2H}, the large cross 
section $\times$ branching fraction of the $gg \to H \to WW^{*}$ mode 
extends the 3-$\sigma$ detection sensitivity of Run 2$^{\mathit{bis}}$ into the 
region $145\gevcc \ltap M_{H} \ltap 180\gevcc$.  This is an extremely 
exciting opportunity, and it is important that the $WW^{*}$ proposal 
receive independent critical analysis.  

\subsection*{Constraints on the Higgs-Boson Mass}
The electroweak theory itself provides reason to expect that 
discoveries will not end with the Higgs boson.  Outside a 
narrow window of Higgs-boson masses, the electroweak theory cannot be 
complete.\footnote{The origin of the mass bounds is reviewed in 
Ref.\ \cite{granada}, where complete references will be found.}

Scalar field theories make sense on all energy scales only if they are 
noninteracting, or ``trivial.'' For any given Higgs-boson mass, there 
is a maximum energy scale $\Lambda^\star$ at which the theory ceases 
to make sense.  Equivalently, if the theory is to be valid up to a 
certain scale $\Lambda^{\star}$, that implies an upper bound on the 
Higgs-boson mass.  The description of the Higgs boson as an elementary 
scalar is at best an effective theory, valid over a finite range of 
energies.  A perturbative analysis identifies
\begin{equation}
\Lambda^{\star} \le M_H\exp{\left(\frac{4\pi^2v^2}{3M_H^2}\right)}\;,
\end{equation}
where $v=(G_F\sqrt{2})^{-1/2}\approx 246~\hbox{GeV}$ is $\sqrt{2}$ 
times the vacuum expectation value of the Higgs field.

This perturbative analysis breaks down when the Higgs-boson mass 
approaches $1\tevcc$ and the interactions become strong.  
Lattice analyses indicate that, for the theory to describe 
physics to an accuracy of a few percent up 
to a few TeV, the mass of the Higgs boson can be no more than about 
$710\pm 60\gevcc$.  Another way of putting this result is that, if 
the elementary Higgs boson takes on the largest mass allowed by 
perturbative unitarity arguments, the electroweak theory will be living 
on the brink of instability.

A lower bound is obtained by 
computing the first quantum corrections to the classical potential
$V(\varphi^{\dagger}\varphi) = \mu^{2}(\varphi^{\dagger}\varphi) + 
\abs{\lambda}(\varphi^{\dagger}\varphi)^{2}$. Requiring that 
$\vev{\varphi}\neq 0$ be an absolute minimum of the one-loop 
potential up to a scale $\Lambda$ yields the vacuum-stability condition 
\begin{equation}
	M_H^2 > \frac{3G_F\sqrt{2}}{8\pi^{2}}(2M_W^4+M_Z^4-4m_{t}^{4})
	\log(\Lambda^{2}/v^{2}) \; .
\end{equation}
\begin{figure}[tb]
	\centerline{\BoxedEPSF{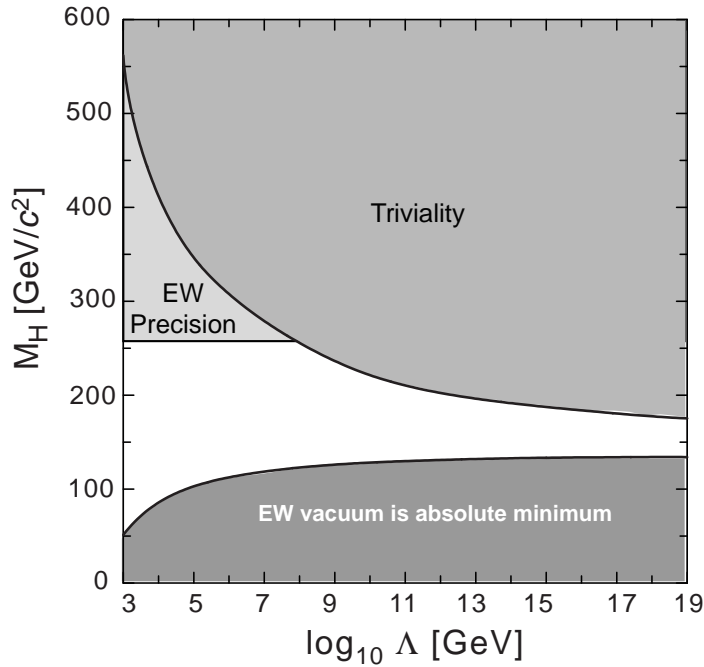  scaled 700}}
	\vspace*{6pt}
	\caption{Bounds on the Higgs-boson mass that follow from requirements
	that the electroweak theory be consistent up to the energy scale
	$\Lambda$.  The upper bound follows from triviality conditions; the
	lower bound follows from the requirement that the minimum of $V$ occur
	for $\varphi \ne 0$.  Also shown is the range of masses excluded at
	the 95\%\ confidence level by precision measurements.}
	\protect\label{fig:Hbds}
\end{figure}

The upper and lower bounds plotted in Figure \ref{fig:Hbds} are the results of 
full two-loop calculations \cite{2loopvacstab}.  There I have also 
indicated the upper bound on $M_{H}$ derived from precision 
electroweak measurements \textit{in the framework of the standard electroweak 
theory.}  If the Higgs boson is relatively light---which would itself require 
explana\-tion---then the theory can be self-consistent up to 
very high energies.  If the electroweak theory is to make sense all the 
way up to a unification scale $\Lambda^\star = 10^{16}~\hbox{GeV}$, then 
the Higgs-boson mass must lie in the interval $145\gevcc \ltap M_{W} \ltap 170
\gevcc$.

\subsection*{The Minimal Supersymmetric Standard Model}
\begin{figure}[t!]
	\centerline{\BoxedEPSF{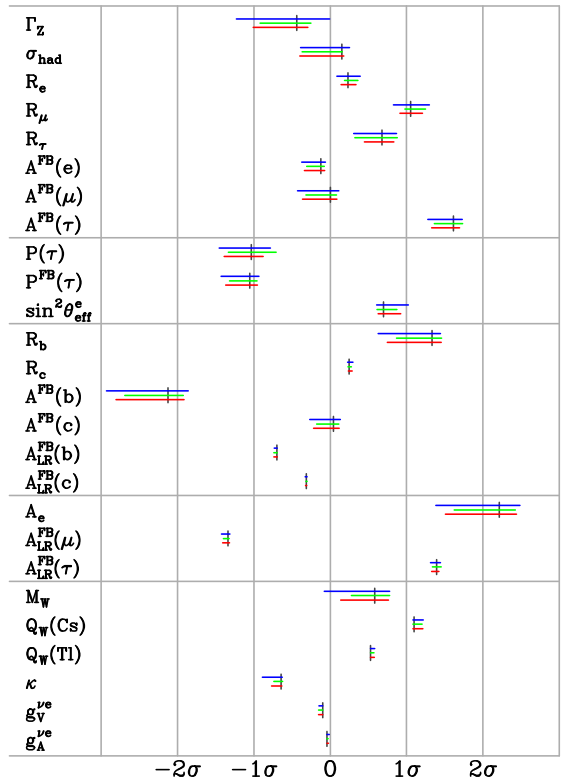  scaled 600}}
	\vspace*{6pt}
\caption{The range of best fit predictions of precision observables in
the supergravity model (upper horizontal lines), the $\mathbf{5} \oplus \mathbf{5^{*}}$
gauge-mediated model (middle lines), the 
$\mathbf{10} \oplus \mathbf{10^{*}}$ gauge-mediated model
(lower lines), and in the standard model at its global best fit value
(vertical lines), in units of standard deviation, from Ref.\ {\protect \cite{Erler:1998ur}}.}
	\protect\label{fig:allobs}
\end{figure}

One of the best phenomenological motivations for supersymmetry on the 
\onetev\ is that the minimal supersymmetric extension of the standard 
model so closely approximates the standard model itself.  A nice 
illustration of the small differences between predictions of 
supersymmetric models and the standard model is the compilation of 
pulls prepared by Erler and Pierce \cite{Erler:1998ur}, which is shown 
in Figure \ref{fig:allobs}.  This is a 
nontrivial property of new physics beyond the standard model, and a 
requirement urged on us by the unbroken quantitative success of the 
established theory.  On the aesthetic side, supersymmetry 
is the maximal---indeed, unique---extension of Poincar\'{e} invariance.  
It also offers a path to the incorporation of gravity, since local 
supersymmetry leads directly to supergravity.  As a practical matter, 
supersymmetry on the \onetev\ offers a solution to the naturalness 
problem, and allows a fundamental scalar to exist at low energies.

Many recent papers investigate the constraints that precision
electroweak measurements might place on the spectrum of superpartners. 
As a representative example, let me point to the analysis of Cho and
Hagiwara\cite{kaoru} already cited for the standard-model fit.  Their
best overall fit, including the particles of the minimal
supersymmetric standard model, suggests that the lightest readily
observable superpartners should be (wino-like) charginos, with masses
in the neighborhood of $100\gevcc$.  The mass of the lightest
neutralino, $\tilde{\chi}_{1}^{0}$, is about $50\gevcc$, while most of
the doublet squarks weigh more than about $200\gevcc$.  The sleptons
are relatively heavy, the lightest being the $\tilde{\tau}_{1}$, in
the range $134\hbox{ - }169\gevcc$.  For the moment, I think it is
prudent to take statements of this kind as mildly suggestive.  As we 
learn more about the mass of the Higgs boson, and constrain the 
standard-model contributions to electroweak observables still further, 
our ability to derive definite expectations for physics beyond the 
standard model should grow.

\begin{figure}[t!]
	\centerline{\BoxedEPSF{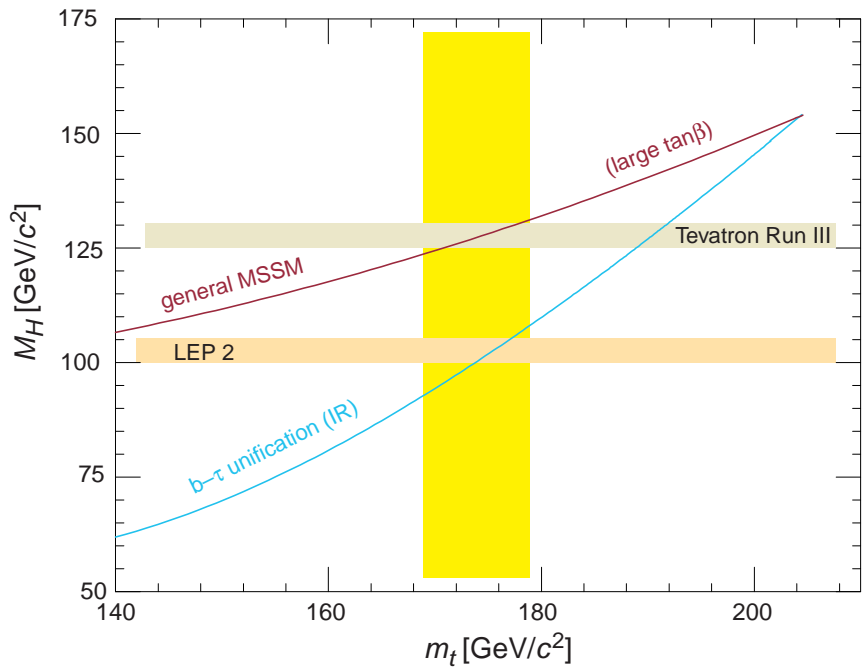  scaled 600}}
	\vspace*{6pt}
	\caption{Upper bounds on the mass of the lightest Higgs boson, as a 
	function of the top-quark mass, in two variants of the minimal 
	supersymmetric standard model.  The upper curve refers to a general 
	MSSM, in the large-$\tan\beta$ limit; the lower curve corresponds to 
	an infrared-fixed-point scenario with $b$-$\tau$ unification, from 
	Ref.\ {\protect \cite{Carena:1995bx}}.}
	\protect\label{fig:MHmt}
\end{figure}
Because the minimal supersymmetric standard model (MSSM) implies 
\textit{upper bounds} on the mass of the lightest scalar $h^{0}$, it 
sets attractive targets for experiment.  Two such upper bounds are 
shown as functions of the top-quark mass in Figure \ref{fig:MHmt}.  
The large-$\tan\beta$ limit of a general MSSM yields the upper curve; 
an infrared-fixed-point scheme with $b$-$\tau$ unification produces 
an upper bound characterized by the lower curve.  The vertical band 
shows the current information on $m_{t}$.  The projected 
sensitivity of LEP~2 experiments \cite{saulan} covers the full range of 
lightest-Higgs masses that occur in the infrared-fixed-point scheme.  
The sensitivity promised by Run 2$^{\mathit{bis}}$ of the Tevatron 
\cite{hobbs} gives full 
coverage of $h^{0}$ masses in the MSSM.  These are very intriguing 
experimental possibilities.  For further discussion, consult the LEP~2 
Yellow Book \cite{Carena:1996bj} and the forthcoming Proceedings of the Tevatron 
Supersymmetry / Higgs Workshop.\footnote{Current drafts are available 
at \textsf{http://fnth37.fnal.gov/susy.html.}}

\subsection*{How Much Can We Trust the Standard-Model Analysis \\ of 
Precision Electroweak Data to Bound $M_{H}$?}
As suggestive as we may find the standard-model fits 
to precision electroweak data that tell us the Higgs boson is light, 
we know that the standard model must break down at a nearby energy 
scale if the Higgs boson is indeed light.  We must ask, with some 
urgency, how the occurrence of new physics near the \onetev\ would 
modify our expectations for $M_{H}$.  Hall and Kolda \cite{hallk} have 
given one provocative illustration.  When they include 
higher-dimension operators that could arise if the dimensionality of 
spacetime is greater than $3\oplus 1$, their fit to the electroweak 
observables leads to no constraint on the Higgs-boson mass: \textit{the 
constraints of the standard-model analysis simply evaporate.}  That 
is a dramatic result, but we do not know whether the situation they 
envisage actually arises in a real theory, and whether the vacuum of 
that theory indeed breaks the electroweak symmetry.
In a similar spirit, Rizzo and Wells \cite{tomjim} have shown that if 
the Higgs boson is trapped on a $3\oplus 1$-dimensional wall with the 
fermions, large Higgs masses (up to $500\gevcc$) and relatively light 
Ka\l uza-Klein mass scales provide a good fit to 
precision data.

\begin{figure}[tb] 
\centerline{\BoxedEPSF{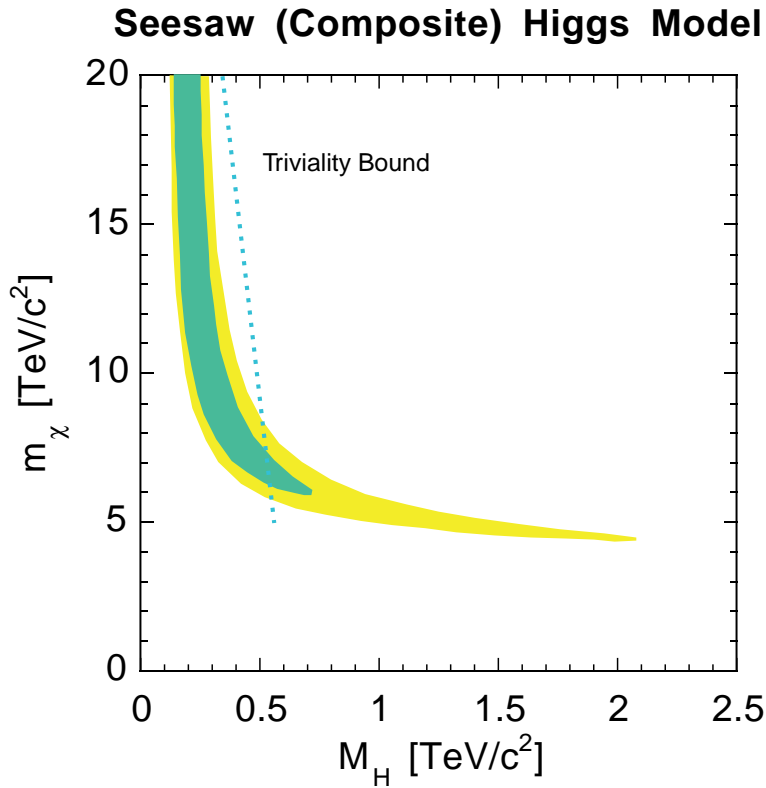 scaled 600}}
\vspace{10pt}
\caption{Precision electroweak constraints (at 68\% and 90\% CL) on 
the Higgs-boson mass and heavy-fermion mass in a top-quark seesaw 
model of a composite Higgs boson, {\protect\cite{tseesaw}}.  The 
triviality bound is from the work of Chivukula and Evans, Ref.\ 
{\protect\cite{rscne}}.}
\label{fig:topcolor}
\end{figure}

More apposite, because it does correspond to a fully elaborated 
theory, is the topcolor seesaw example recently offered by Collins, 
Grant, and Georgi \cite{tseesaw}.  In a model that includes one 
additional heavy weak-singlet fermion $\chi$, with weak hypercharge 
$Y(\chi) = \cfrac{4}{3}$, they find that the mass of the Higgs boson 
can exceed $300\gevcc$ (the $\gtap 90\%$ CL upper bound from 
standard-model analyses) for a heavy-fermion mass in the range 
$5\tevcc \ltap m_{\chi} \ltap 7\tevcc$.  The 1-$\sigma$ and 90\% CL 
allowed regions in the $(m_{\chi},M_{H})$ plane are shown in Figure 
\ref{fig:topcolor}.\footnote{A second topcolor seesaw model based on 
two heavy fermions and two Higgs doublets leads to heavy scalars 
$h^{0}, H^{0}, H^{\pm}$, but a \textit{light pseudoscalar}---quite 
different from the spin-0 spectrum of the MSSM.}

We need more examples based on plausible---or at least 
provocative---extensions of the standard model to help us understand 
how seriously to take the standard-model clues, and to help us think 
about what it might mean if the standard-model hopes are dashed.

\subsection*{Looking for Trouble}
Suppose, in the face of the spectacular successes of the electroweak 
theory, we go looking for trouble.  Where might we find it?  The 
great mass of the top quark gives rise to the theoretical suspicion that 
anomalies are most likely to show themselves in the third generation 
of quarks and leptons.  As it happens, the largest pull in precision 
measurements on the $Z^{0}$ pole involves $b$ quarks.  The 
forward-backward asymmetry for $b\bar{b}$ events measured at LEP and 
the left-right forward-backward asymmetry for $b\bar{b}$ events measured 
at SLD indicate a three-standard-deviation difference from the 
standard model for 
\begin{equation}
	A_{b} = \frac{L_{b}^{2} - R_{b}^{2}}{L_{b}^{2} + R_{b}^{2}} \; ,
	\label{eq:Abdef}
\end{equation}
where $L_{b}$ and $R_{b}$ are the left-handed and right-handed chiral 
couplings of the $Z$ to $b$ quarks.  
Pursuing this line, we arrive at the suggestion that $\delta 
R_{b}/R_{b}^{\mathrm{theory}} \approx -40\%$, whereas $\delta 
L_{b}/L_{b}^{\mathrm{theory}} \approx 1\%$.  Therefore,
if we want to find deviations from standard-model predictions, we 
should try to isolate effects of the right-handed $b$ 
coupling \cite{Marciano:1999ia}.  If this anomaly is real, we might 
expect to observe flavor-changing neutral-current transitions $b \to 
s$, $b \to d$, and $s \to d$.

Bennett and Wieman (Boulder) have reported a new determination of the 
weak charge of Cesium by measuring the transition polarizability for the 
6S-7S transition \cite{benwie}.  The new value,
\begin{equation}
	Q_W(\textrm{Cs})= -72.06 \pm 0.28\hbox{ (expt)} \pm 0.34\hbox{ 
	(theory),} 
	\label{eq:wieman}
\end{equation}
represents a sevenfold improvement in the experimental error and a 
significant reduction in the theoretical uncertainty.  It lies about
2.5 standard deviations above the prediction of the standard model.  We 
are left with the traditional situation in which elegant measurements 
of parity nonconservation in atoms are on the edge of incompatibility 
with the standard model.

A number of authors \cite{erlang} have noted that the discrepancy in
the weak charge $Q_{W}$ and a 2-$\sigma$ anomaly\footnote{In Erler \&
Langacker's fit, for example, the number of light neutrino species
inferred from the invisible width of the $Z^{0}$ is $N_{\nu} = 2.985
\pm 0.008 = 3 - 2\sigma$.} in the total width of the $Z^{0}$ can be
reduced by introducing a $Z^{\prime}$ boson with a mass of about
$800\gevcc$.  The additional neutral gauge boson resembles the
$Z_{\chi}$ familiar from unified theories based on the group $E_{6}$.

\subsection*{The Vacuum Energy Problem}
I want to spend a moment to revisit a 
longstanding, but usually unspoken, challenge to the completeness of 
the electroweak theory as we have defined it: the vacuum energy 
problem \cite{Veltman:1975au}.
I do so not only for its intrinsic interest, but also to 
raise the question, ``Which problems of completeness and 
consistency do we worry about at a given moment?''  It is perfectly 
acceptable science---indeed, it is often essential---to put certain 
problems aside, in the expectation that we will return to them at the 
right moment.  What is important is never to forget that the problems 
are there, even if we do not allow them to paralyze us.  

For the usual Higgs potential, 
$V(\varphi^{\dagger}\varphi) = \mu^{2}(\varphi^{\dagger}\varphi) + 
\abs{\lambda}(\varphi^{\dagger}\varphi)^{2}$, the value of 
the potential at the minimum is
\begin{equation}
    V(\vev{\varphi^{\dagger}\varphi}) = \frac{\mu^{2}v^{2}}{4} = 
    - \frac{\abs{\lambda}v^{4}}{4} < 0.
    \label{minpot}
\end{equation}
Identifying $M_{H}^{2} = -2\mu^{2}$, we see that the Higgs potential 
contributes a field-independent constant term,
\begin{equation}
    \varrho_{H} \equiv \frac{M_{H}^{2}v^{2}}{8}.
    \label{eq:rhoH}
\end{equation}
I have chosen the notation $\varrho_{H}$ because the constant term in the 
Lagrangian plays the role of a vacuum energy density.  When we 
consider gravitation, adding a vacuum energy density 
$\varrho_{\mathrm{vac}}$ is equivalent to adding a cosmological constant 
term to Einstein's equation.  Although recent observations 
\footnote{For a cogent summary of current knowledge of the 
cosmological parameters, including evidence for a cosmological 
constant, see Ref.\ \cite{cosconst}.} raise the intriguing 
possibility that the cosmological constant may be different from zero, 
the essential observational fact is that the vacuum energy density 
must be very tiny indeed,\footnote{For a useful summary of 
gravitational theory, see the essay by T.  d'Amour in \S14 of the 1998 
\textit{Review of Particle Physics,} Ref.\ \cite{pdg}.}
\begin{equation}
    \varrho_{\mathrm{vac}} \ltap 10^{-46}\gev^{4}\; .
    \label{eq:rhovaclim}
\end{equation}
Therein lies the puzzle: if we take
$v = (G_F\sqrt{2})^{-\frac{1}{2}}  \approx 246\gev$  
and insert the current experimental lower bound 
\cite{saulan}
$M_{H} \gtap 105\gevcc$ into \eqn{eq:rhoH}, we find that the 
contribution of the Higgs field to the vacuum energy density is
\begin{equation}
    \varrho_{H} \gtap 8 \times 10^{7}\gev^{4},
    \label{eq:rhoHval}
\end{equation}
some 54 orders of magnitude larger than the upper bound inferred from 
the cosmological constant.

What are we to make of this mismatch?  The fact that $\varrho_{H} \gg 
\varrho_{\mathrm{vac}}$ means that the smallness of the cosmological 
constant needs to be explained.  In a unified theory of the strong, 
weak, and electromagnetic interactions, other (heavy!) Higgs fields 
have nonzero vacuum expectation values that may give rise to still 
greater mismatches.  At a fundamental level, we can therefore conclude 
that a spontaneously broken gauge theory of the strong, weak, and 
electromagnetic interactions---or merely of the electroweak 
interactions---cannot be complete.  Either we must find a separate 
principle to zero the vacuum energy density of the Higgs field, or 
we may suppose that a proper quantum theory of gravity, in combination 
with the other interactions, will resolve the puzzle of the 
cosmological constant.  The vacuum energy problem must be an important 
clue.  But to what?

%\clearpage
\section*{The Problems of Mass, and of Mass Scales}
Electroweak symmetry breaking sets the 
values of the $W$- and $Z$-boson masses.  At tree level in the 
electroweak theory, we have
\begin{eqnarray}
    M_{W}^{2} & = & g^{2}v^{2}/2 = 
    \pi\alpha/G_{F}\sqrt{2}\sin^{2}\theta_{W} ,
    \label{eq:Wmass}  \\
    M_{Z}^{2} & = & M_{W}^{2}/\cos^{2}\theta_{W} ,
    \label{eq:Zmass}  
\end{eqnarray}
where the electroweak scale is $v = (G_{F}\sqrt{2})^{-\frac{1}{2}}
\approx 246\gev$.  But the electroweak scale is not the only scale.  
It seems certain that we must also consider the Planck scale, derived 
from the strength of Newton's constant,
\begin{equation}
    M_{\mathrm{Planck}} =  (\hbar c/G_{\mathrm{Newton}})^{\frac{1}{2}} \approx 1.22 \times 10^{19}\gev \; .
    \label{eq:planck}
\end{equation}
It is also probable that we must take account of the $SU(3)_{c}\otimes 
SU(2)_{L}\otimes U(1)_{Y}$ unification scale around 
$10^{15\mathrm{-}16}\gev$, and there may well be a distinct 
flavor scale.  The existence of other scales is behind the famous 
problem of the Higgs scalar mass: how to keep the distant scales from 
mixing in the face of quantum corrections, or how to stabilize the 
mass of the Higgs boson on the electroweak scale.

It is because $G_{\mathrm{Newton}}$ is so small (or because 
$M_{\mathrm{Planck}}$ is so large) that we normally consider 
gravitation irrelevant for particle physics.  The 
graviton-quark-antiquark coupling is generically $\sim 
E/M_{\mathrm{Planck}}$, so it is easy to make a dimensional estimate 
of the branching fraction for a gravitationally mediated rare kaon 
decay: $B(K_{L} \to \pi^{0}G) \sim (M_{K}/M_{\mathrm{Planck}})^{2} \sim 
10^{-38}$, which is truly negligible!

We know from the electroweak theory alone that the 1-TeV scale is 
special.  Partial-wave unitarity applied to gauge-boson scattering 
tells us that unless the Higgs-boson mass respects
\begin{equation}
    M_{H}^{2} < \frac{8\pi\sqrt{2}}{3G_{F}} \approx (1\tevcc)^{2} \; ,
    \label{eq:higgsu}
\end{equation}
new physics is to be found on the 1-TeV scale \cite{lqt}.  To stabilize the
Higgs-boson mass against uncontrolled quantum corrections, and to
resolve the mass-hierarchy problem, we consider electroweak physics
beyond the standard model.  The most promising approaches are to 
generalize $SU(3)_{c}\otimes SU(2)_{L}\otimes U(1)_{Y}$ to a theory 
with a composite Higgs boson in which the electroweak symmetry is
broken dynamically (technicolor and related theories) or to a
supersymmetric standard model.

Let us look a little further at the problem of fermion 
masses.\footnote{For an overview of the standard-model approach to 
fermion mass, see Ref.\ \cite{fwfm}.}  In the 
electroweak theory, the value of each quark or charged-lepton mass is 
set by a new, unknown, Yukawa coupling.  We define the left-handed 
doublets and right-handed singlets
\begin{equation}
    {\mathsf{L}}_{e} = \left( 
    \begin{array}{c}
        \nu_{e}  \\
        e
    \end{array}
    \right)_{L} \; , \qquad {\mathsf{R}}_{e} \equiv e_{R} \;;
    \qquad
    {\mathsf{L}}_{q} = \left( 
    \begin{array}{c}
        u  \\
        d
    \end{array}
    \right)_{L} \; , \qquad {\mathsf{R}}_{u} \equiv u_{R}\; ,
    \quad {\mathsf{R}}_{d} \equiv d_{R}\; .
    \label{eq:elec}
\end{equation}
Then the electron's Yukawa term in the electroweak Lagrangian is
\begin{equation}
    {\mathcal{L}}_{\mathrm{Yukawa}}^{(e)} = - 
    \zeta_{e}[\bar{\mathsf{R}}_{e}(\varphi^{\dagger}{\mathsf{L}}_{e}) + 
    (\bar{\mathsf{L}}_{e}\varphi){\mathsf{R}}_{e}] \; ,
    \label{eq:eYuk}
\end{equation}
where $\varphi$ is the Higgs field, so that the electron mass is
$m_{e} = \zeta_{e}v/\sqrt{2}$.  Similar expressions obtain for the 
quark Yukawa couplings.  Inasmuch as we do not know how to
calculate the fermion Yukawa couplings $\zeta_{f}$, I believe that
\textit{we should consider the sources of all fermion masses as
physics beyond the standard model.}  

The values of the Yukawa couplings are vastly different for
different fermions: for the top quark, $\zeta_{t} \approx 1$, for the
electron $\zeta_{e} \approx 3\times 10^{-6}$, and if the neutrinos
have Dirac masses, presumably $\zeta_{\nu} \approx
10^{-10}$.\footnote{I am quoting the values of the Yukawa couplings at
a low scale typical of the masses themselves.} What accounts for the
range and values of the Yukawa couplings?  Our best hope until now has
been the suggestion from unified theories that the pattern of fermion
masses simplifies on high scales.  The classic intriguing prediction
of the $SU(5)$ unified theory involves the masses of the $b$ quark and
the $\tau$ lepton, which are degenerate at the unification point for a
simple pattern of spontaneous symmetry breaking.  The different
running of the quark and lepton masses to low scales then leads to the
prediction $m_{b} \approx 3 m_{\tau}$, in suggestive agreement with
what we know from experiment.

\begin{figure}[tb] 
\centerline{\BoxedEPSF{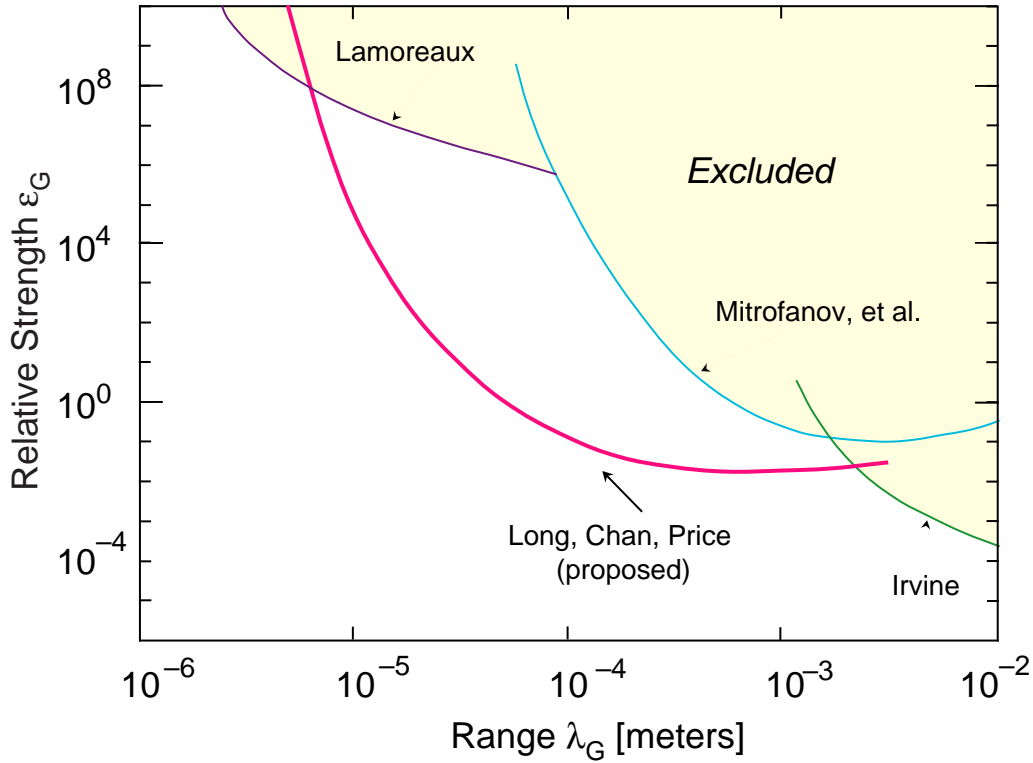 scaled 750}}
\vspace{10pt}
\caption{Experimental limits on the strength $\varepsilon_{\mathrm{G}}$ 
(relative to gravity) versus the range $\lambda_{\mathrm{G}}$ of a 
new long-range force, together with the anticipated sensitivity of a 
new experiment based on small mechanical resonators {\protect\cite{price}}.}
\label{fig:nonNgrav}
\end{figure}

The conventional approach to new physics has been to extend the 
standard model to understand why the electroweak scale (and the mass 
of the Higgs boson) is so much smaller than the Planck scale.  A novel 
approach that has been developed over the past two years is instead to 
\textit{change gravity} to understand why the Planck scale is so much 
greater than the electroweak scale \cite{EDbiblio}.  Now, experiment 
tells us that gravitation closely follows the Newtonian force law down 
to distances on the order of $1\mm$.  Let us parameterize deviations 
from a $1/r$ gravitational potential in terms of a relative strength 
$\varepsilon_{\mathrm{G}}$ and a range $\lambda_{\mathrm{G}}$, so that
\begin{equation}
V(r) = - \int dr_{1}\int dr_{2} 
\frac{G_{\mathrm{Newton}}\rho(r_{1})\rho(r_{2})}{r_{12}} \left[ 1+ 
\varepsilon_{\mathrm{G}}\exp(-r_{12}/\lambda_{\mathrm{G}}) \right]\; ,	
\end{equation}
where $\rho(r_{i})$ is the mass density of object $i$ and $r_{12}$ is 
the separation between bodies 1 and 2.
Elegant experiments that study details of Casimir and Van der Waals 
forces imply bounds on anomalous gravitational interactions, as shown 
in Figure \ref{fig:nonNgrav}.  Below about a millimeter, the 
constraints on deviations from Newton's inverse-square force law deteriorate 
rapidly, so nothing prevents us from considering changes to gravity 
even on a small but macroscopic scale.

For its internal consistency, string theory requires an additional six 
or seven space dimensions, beyond the $3+1$ dimensions of everyday 
experience.  Until recently it has been presumed that the extra 
dimensions must be compactified on the Planck scale, with a
compactification radius $R_{\mathrm{unobserved}} \approx
1/M_{\mathrm{Planck}} \approx 1.6 \times 10^{-35}\m$.  The new wrinkle 
is to consider that the $SU(3)_{c}\otimes SU(2)_{L}\otimes U(1)_{Y}$
standard-model gauge fields, plus needed extensions, reside on 
$3+1$-dimensional branes, not in the extra dimensions, but that 
gravity can propagate into the extra dimensions.

\begin{figure}[tb] 
\centerline{\BoxedEPSF{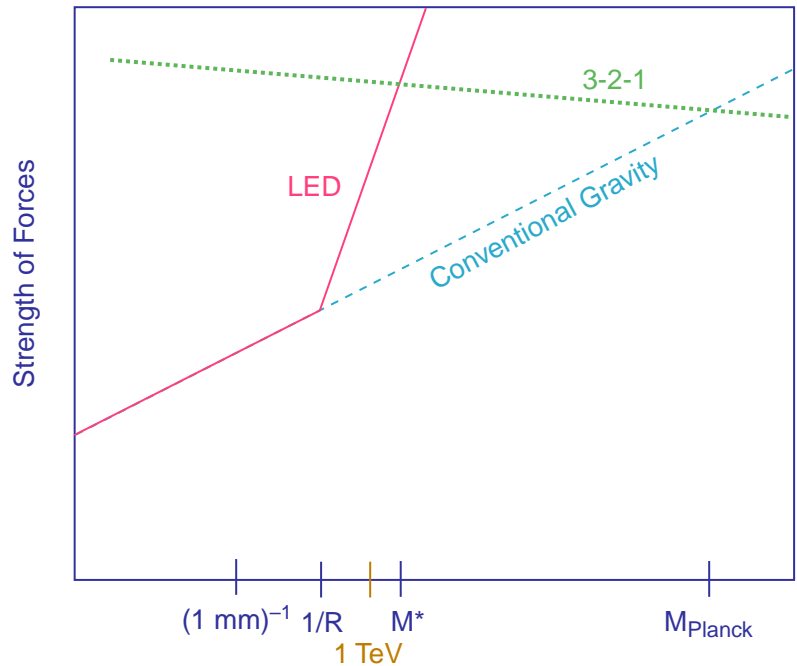 scaled 750}}
\vspace{10pt}
\caption{One of these extrapolations (at least!) is false.}
\label{fig:false}
\end{figure}

How does this hypothesis change the picture?  The dimensional 
analysis (Gauss's law, if you like) that relates Newton's constant to 
the Planck scale changes.  If gravity propagates in $n$ extra 
dimensions with radius $R$, then
\begin{equation}
    G_{\mathrm{Newton}} \sim M_{\mathrm{Planck}}^{-2} \sim M^{\star\,-n-2}R^{-n}\; ,
    \label{eq:gauss}
\end{equation}
where $M^{\star}$ is gravity's true scale.  Notice that if we boldly 
take $M^{\star}$ to be as small as $1\tevcc$ (which other constraints 
may not allow), then the radius of the extra 
dimensions is required to be smaller than about $1\mm$, for $n \ge 
2$.  If we use the four-dimensional force law to extrapolate the 
strength of gravity from low energies to high, we find that gravity 
becomes as strong as the other forces on the Planck scale, as shown 
by the dashed line in Figure \ref{fig:false}.  If the force law 
changes at an energy $1/R$, as the large-extra-dimensions scenario 
suggests, then the forces are unified at an energy $M^{\star}$, as 
shown by the solid line in Figure \ref{fig:false}.
What we know as the Planck scale is then a mirage that results 
from a false extrapolation: treating gravity as four-dimensional down 
to arbitrarily small distances, when in fact---or at least in this 
particular fiction---gravity propagates in $3+n$ spatial dimensions.  
The Planck mass is an artifact, given by $M_{\mathrm{Planck}} = 
M^{\star}(M^{\star}R)^{n/2}$. 

Although the idea that extra dimensions are just around the 
corner---either on the submillimeter scale or on the TeV scale---is 
preposterous, it is not ruled out by observations.  For that reason 
alone, we should entertain ourselves by entertaining the 
consequences.  Many authors have considered the gravitational 
excitation of a tower of Ka\l uza--Klein modes in the extra 
dimensions, which would give rise to a missing (transverse) energy 
signature in collider experiments \cite{smaria}.  We call these excitations 
\textit{provatons,} after the Greek word for a sheep in a 
flock.
%%%%%%%%%%%%%%%%%%%%%%%%%%%%%%%%%%%%%%%%%%%%%%%%%%%%%%%%%%%%%%%%%%%%%%
%                                                                    %
%   \footnote{I thank Maria Spiropulu for instructing me in the      %
%   difference between $\pi\rho\acute{o}\beta\alpha\tau o\nu$ and    %
%   $\alpha\rho\nu\grave{\iota}$.}                                   %
%                                                                    %
%%%%%%%%%%%%%%%%%%%%%%%%%%%%%%%%%%%%%%%%%%%%%%%%%%%%%%%%%%%%%%%%%%%%%%

\begin{figure}[tb] 
\centerline{\BoxedEPSF{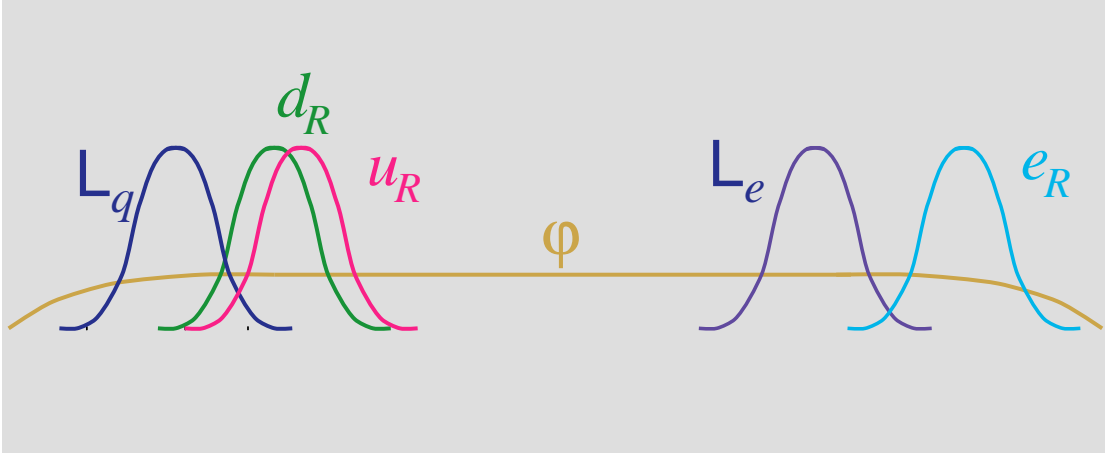 scaled 750}}
\vspace{10pt}
\caption{The locations in the brane of wave packets that represent 
different fermion species.}
\label{fig:braneferm}
\end{figure}

``Large'' extra dimensions present us with new ways to think about the 
exponential range of Yukawa couplings.  Arkani-Hamed, Schmaltz, and 
Mirabelli \cite{martin} have speculated that if the standard-model 
brane has a small thickness, the wave packets representing different 
fermion species might have different locations within the extra 
dimension, as depicted in Figure \ref{fig:braneferm}.  On this 
picture, the Yukawa couplings measure the overlap in the extra 
dimensions of the left-handed and right-handed wave packets and the 
Higgs field, presumed pervasive.  Exponentially large differences 
might then arise from small offsets in the new coordinate(s).  True or 
not, it is a completely different way of looking at an important 
problem.  Speculations of this sort give a new force to the useful 
metaphor of particle accelerators as giant microscopes.  We may soon 
be able to examine Nature on such a small scale that we uncover not 
only fine new features of the familiar ground, but also reveal 
hitherto unknown dimensions.

\section*{QCD and the Electroweak Theory}
Because QCD is asymptotically free and becomes strong at low energies, 
it has a rich phase structure.  It is worth asking whether the phases 
that arise in QCD hold lessons for electroweak symmetry breaking.  We 
already know of one case for which the answer is Yes.

Our conception of electroweak symmetry breaking is modeled upon our understanding of the superconducting 
phase transition.  The Higgs 
mechanism of the standard model is the relativistic generalization of 
the Ginzburg-Landau description of the superconducting phase 
transi\-tion.  The macroscopic order parameter of the Ginzburg-Landau 
phenomenology corresponds to the wave function of superconducting 
charge carriers, which acquires a nonzero vacuum expectation value in 
the superconducting state and generates a photon mass that explains 
the Meissner effect.  The microscopic Bardeen-Cooper-Schrieffer 
theory identifies the dynamical origin of the order 
parameter with the formation of correlated states of elementary fermions, 
the Cooper pairs of electrons.  Can we find a dynamical mechanism for 
electroweak symmetry breaking that corresponds to the BCS theory?

The elementary fermions---electrons---and 
gauge interactions---QED---needed to generate the scalar bound states are 
already present in the case of superconductivity. Could a scheme
 of similar economy account
for the transition that hides the electroweak symmetry?
Consider an $SU(3)_c\otimes SU(2)_L\otimes U(1)_Y$ theory of massless up and 
down quarks. Because the strong interaction is strong, and the electroweak 
interaction is feeble, we may treat the $SU(2)_L\otimes U(1)_Y$ 
interaction as a perturbation. For vanishing quark masses, QCD has an exact 
$SU(2)_L\otimes SU(2)_R$ chiral symmetry. At an energy scale 
$\sim\Lambda_{\mathrm{QCD}},$ the strong interactions become strong, fermion 
condensates appear, and the chiral symmetry is spontaneously broken
to the familiar flavor (isospin) symmetry:
\begin{equation}
	SU(2)_L\otimes SU(2)_R \to SU(2)_V\;\; .
\end{equation}
 Three Goldstone bosons appear, one for 
each broken generator of the original chiral invariance. These were 
identified by Nambu~\cite{20} as three massless pions.

The broken generators are three axial currents whose couplings to pions are 
measured by the pion decay constant $f_\pi = 93\mev$. When we turn on the 
$SU(2)_L\otimes U(1)_Y$ electroweak interaction, the electroweak gauge 
bosons couple to the axial currents and acquire masses of order $\sim 
gf_\pi$. The mass-squared matrix,
\begin{equation}
	{\mathcal{M}}^{2} = \left(
		\begin{array}{cccc}
		g^{2} & 0 & 0 & 0  \\
		0 & g^{2} & 0 & 0  \\
		0 & 0 & g^{2} & gg^{\prime}  \\
		0 & 0 & gg^{\prime} & g^{\prime2}
	\end{array}
		 \right) \frac{f_{\pi}^{2}}{4} \; ,
	\label{eq:csbm2}
\end{equation}
(where the rows and columns correspond to the weak-isospin gauge 
bosons $W^{+}$, $W^{-}$, $W_{3}$, and the weak-hypercharge gauge 
boson $\mathcal{A}$) has the same structure as the mass-squared matrix 
for gauge bosons in the standard electroweak theory.  Diagonalizing 
the matrix \eqn{eq:csbm2}, we find that $M_{W}^{2} = 
g^{2}f_{\pi}^{2}/4$ and $M_{Z}^{2} = 
(g^{2}+g^{\prime2})f_{\pi}^{2}/4$, so that
\begin{equation}
	\frac{M_{Z}^{2}}{M_{W}^{2}} = \frac{(g^{2}+g^{\prime2})}{g^{2}} = 
	\frac{1}{\cos^{2}\theta_{W}}\; .
	\label{eq:wzrat}
\end{equation}
The photon emerges massless.

The massless pions thus disappear from the physical spectrum, 
having become the longitudinal components of the weak gauge bosons. 
Unfortunately, the mass acquired by the 
intermediate bosons is far smaller than required for a successful 
low-energy phenomenology; $M_W\approx 30\mevcc$ and $M_{Z} \approx 
34\mevcc$, about $\cfrac{1}{2650} \times$ the true masses \cite{21}.

If only we didn't know $f_{\pi}$!  The idea of replacing the 
elementary Higgs boson with a fermion-antifermion condensate is too 
good to abandon without a struggle \cite{Chivukula:1996uy}. Perhaps a 
more general formulation of the chiral-symmetry--breaking idea has 
merit.  We replace the massless up and down quarks with new 
``technifermions'' and replace QCD with a new ``technicolor'' gauge 
interaction \cite{22,23}. We choose the scale of the interaction---the 
analogue of $\Lambda_{\mathrm{QCD}}$---so that
\begin{equation}
    f_{\pi} \to F_{\pi} = v = (G_{F}\sqrt{2})^{-1/2}\; .
    \label{eq:bigfpi}
\end{equation}
Repeating the analysis we have just made for QCD, we predict the 
correct (tree-level) masses for $W^{\pm}$, $Z^{0}$, and the photon.
By analogy with the superconducting phase transition, the 
dynamics of the fundamental technicolor gauge interactions among 
technifermions generate scalar bound states, and these play the role 
of the Higgs fields.

Technicolor shows how the generation of intermediate boson masses 
could arise from strong dynamics (though not the strong dynamics of 
QCD), without fundamental scalars or unnatural adjustments of 
parameters.  It thus provides an elegant solution to the naturalness 
problem of the Standard Model.  However, it has a major deficiency: it 
offers no explanation for the origin of quark and lepton masses, 
because no Yukawa couplings are generated between Higgs fields and 
quarks or leptons \cite{etc}.  Consequently, technicolor serves as a reminder 
that there are two problems of mass: explaining the masses of the 
gauge bosons, which demands an understanding of electroweak symmetry 
breaking; and accounting for the quark and lepton masses, which 
requires not only an understanding of electroweak symmetry breaking 
but also a theory of the Yukawa couplings that set the scale of 
fermion masses in the standard model.  We can be confident that the 
origin of gauge-boson masses will be understood on the 1-TeV scale.  
We do not know where we will decode the pattern of the Yukawa 
couplings.

Is it possible that other interesting phases of QCD---color 
superconductivity \cite{colorsc,fwhd}, for example---might hold 
lessons for electroweak symmetry breaking under normal or unusual 
conditions?

\section*{Concluding Remarks}
Wonderful opportunities await particle physics over the next decade, 
with the coming of the Large Hadron Collider at CERN to explore the 
1-TeV scale (extending the efforts at LEP and the Tevatron to unravel 
the nature of electroweak symmetry breaking) and many initiatives to 
develop our understanding of the problem of identity: what makes a 
neutrino a neutrino and a top quark a top quark.  Here I have in mind 
the work of the $B$ factories and the Tevatron collider on \textsf{CP} violation 
and the weak interactions of the $b$ quark; the wonderfully sensitive 
experiments at Brookhaven, Fermilab, CERN, and Frascati on \textsf{CP} 
violation and rare decays of kaons; the prospect of definitive 
accelerator experiments on neutrino oscillations and the nature of the 
neutrinos; and a host of new experiments on the sensitivity frontier.  
We might even learn to read experiment for clues about the 
dimensionality of spacetime.  If we are inventive enough, we may be 
able to follow this rich menu with the physics opportunities offered 
by a linear collider and a (muon storage ring) neutrino factory.  I 
expect a remarkable flowering of experimental particle physics, and of 
theoretical physics that engages with experiment.

%\section*{Acknowledgments}

\end{document}